%% file: main.tex
\documentclass[openacc]{rstransa}
\usepackage{hyperref}
\usepackage[capitalise]{cleveref}
\usepackage{bm}
\usepackage{subcaption}

\newcommand{\iap}{CNRS \& Sorbonne Universit\'{e}, Institut d’Astrophysique de Paris (IAP), UMR 7095, 98 bis bd Arago, F-75014 Paris, France}
\newcommand{\jhupha}{Department of Physics and Astronomy, Johns Hopkins University, Baltimore, MD 21218, USA}
\newcommand{\oxford}{Astrophysics, University of Oxford, Denys Wilkinson Building, Keble Road, Oxford, OX1 3RH, UK}

\newcommand{\hyper}[1]{\ensuremath{{}_{2}F_{1}}#1}
\newcommand{\Mpch}{\ensuremath{h^{-1}\,\text{Mpc}}}
\newcommand{\hMpc}{\ensuremath{h\,\text{Mpc}^{-1}}}
\newcommand{\splitatcommas}[1]{%
  \begingroup
  \begingroup\lccode`~=`, \lowercase{\endgroup
    \edef~{\mathchar\the\mathcode`, \penalty0 \noexpand\hspace{0pt plus 1em}}%
  }\mathcode`,="8000 #1%
  \endgroup
}

\newcommand{\camb}{\textsc{camb}}

\newcommand{\operon}{\textsc{operon}}

\newcommand{\bacco}{\textsc{bacco}}
\newcommand{\euclidemu}{\textsc{euclidemulator2}}
\newcommand{\cosmopower}{\textsc{cosmopower}}

\newcommand{\halofit}{\textsc{halofit}}

\newcommand{\syren}{\textsc{syren}}
\newcommand{\jaxcosmo}{\textsc{jax-cosmo}}
\newcommand{\cosmosis}{\textsc{cosmoSIS}}
\newcommand{\eh}{E\&H}
\DeclareMathOperator{\aq}{aq}

\titlehead{Research}

\begin{document}

\include{journals}

\title{Symbolic Emulators for Cosmology: Accelerating Cosmological Analyses Without Sacrificing Precision}

\author{Deaglan J. Bartlett$^{1,2}$ and Shivam Pandey$^{3}$}

\address{$^{1}$\oxford\\
$^{2}$\iap\\
$^{3}$\jhupha}

\subject{Cosmology, Astrophysics, Artificial intelligence}


\corres{Deaglan J. Bartlett\\
\email{deaglan.bartlett@physics.ox.ac.uk}}

\begin{abstract}
In cosmology, emulators play a crucial role by providing fast and accurate predictions of complex physical models, enabling efficient exploration of high-dimensional parameter spaces that would be computationally prohibitive with direct numerical simulations. Symbolic emulators have emerged as promising alternatives to numerical approaches, delivering comparable accuracy with significantly faster evaluation times. While previous symbolic emulators were limited to relatively narrow prior ranges, we expand these to cover the parameter space relevant for current cosmological analyses. 
We introduce approximations to hypergeometric functions used for the $\Lambda$CDM comoving distance and linear growth factor which are accurate to better than 0.001\% and 0.05\%, respectively, for all redshifts and for $\Omega_{\rm m} \in [0.1, 0.5]$.
We show that integrating symbolic emulators into a Dark Energy Survey-like $3\times2$pt analysis produces cosmological constraints consistent with those obtained using standard numerical methods. Our symbolic emulators offer substantial improvements in speed and memory usage, demonstrating their practical potential for scalable, likelihood-based inference.
\end{abstract}

\begin{fmtext}
\end{fmtext}
\maketitle

\section{Introduction}

The aim of any cosmological analysis is to infer parameters which describe our Universe given observed data.
This involves computing a prediction for an observable quantity given some set of cosmological parameters and then comparing these to the observed values. One must repeat this procedure typically thousands of times as one searches through cosmological parameters which could reasonably explain the observations.

These predictions can be prohibitively slow, and thus it has become increasingly common to replace exact simulations with surrogate models which are much faster yet are sufficiently accurate to not bias our predictions. These have conventionally taken the form of numerical machine learning emulators, such as 
\textsc{Aemulus} \cite{Zhai_2019},
\bacco{} \cite{Angulo_2021,Arico_2021,Zennaro_2023},
\textsc{cobra} \cite{Bakx_2024},
\textsc{CosmicEmu} \cite{Lawrence_2017},
\cosmopower{} \cite{SpurioMancini_2022,Piras_2023}.
\textsc{Dark Emulator} \cite{Nishimichi_2019},
\textsc{emuPK} \cite{Mootoovaloo_2022},
\textsc{euclidemulator1} \cite{Knabenhans_2019},
\euclidemu{} \cite{Knabenhans_2021},
\textsc{FofrFittingFunction} \cite{Winther_2019},
\textsc{FrankenEmu} \cite{Heitmann_2009,Heitmann_2014},
\textsc{NGenHalofit} \cite{Smith_2019},
and \textsc{pico} \cite{Pico2,Pico1}.

Although much faster than solving the full simulation, it is hard to interpret what the emulators are doing `under the hood' and, to continue using these into the future, one must rely on the programming languages and packages which they are written with being maintained and commonly used.
As an alternative, one can often approximate the output of these simulations with symbolic models. Such methods were common long before their numerical counterparts, of which the the BBKS \cite{Bardeen_1986} and Eisenstein \& Hu (\eh) \cite{Eisenstein_1998,Eisenstein_1999} approximations for the matter power spectrum are particularly popular. 
Historically, these expressions were obtained through careful, manual, analytic considerations of the outputs of simulations and were able to achieve accuracies of order of a few to tens of percent.
More recently, the supervised machine learning technique of symbolic regression \cite{Kronberger_2024} has been utilised to extract symbolic approximations for cosmological quantities of interest in an automated fashion \cite{Bartlett_2024_linear,Bartlett_2024_syren,Sui_2024,Bayron-Orjuela-Quintana_2023,Orjuela-Quintana_2024,Kammerer_2025}.
Not only are these more interpretable than the numerical emulators, but they can be made just as accurate yet orders of magnitude faster \cite{Bartlett_2024_linear,Bartlett_2024_syren,Sui_2024}.
Moreover, the output is a symbolic equation using common mathematical expressions; these formulae are easily portable and implemented in any language of choice and do not depend on any particular package.
It is possible to analytically differentiate such expressions, which improves efficiency for high-dimensional sampling algorithms. Alternatively, they can be easily imported into programming languages which have automatic differentiation (e.g. \textsc{jax}) to achieve the same goal.

The goals of this work are threefold. First, we aim to extend the prior range of symbolic emulators to match the broader parameter ranges currently used in cosmological analyses, while maintaining the same level of accuracy. 
It is crucial that these emulators are accurate to at least the percent level to be useful for current and upcoming surveys \cite{Taylor_2018}.
These are developed in \cref{sec:wider syren}. Second, we demonstrate that these extended emulators yield consistent cosmological constraints when embedded within a DES Y1-like $3\times2$pt analysis. 
This is the first demonstration that solely symbolic emulators can be used in Stage-III survey applications without causing biases; we leave the study of their ability to be applied to Stage-IV surveys to future work.
Finally, we quantify the practical advantages of the symbolic approach by highlighting improvements in computational speed and memory efficiency. 
This analysis is performed in \cref{sec:cosmo_analysis}.
For simplicity, we focus on $\Lambda$CDM cosmologies in this work. 
It is possible to extend this to other cosmological models, such as with massive neutrinos and $w_0$-$w_a$ with existing tools, as done in previous studies \cite{Sui_2024}. We conclude in \cref{sec:conclusion}.

Throughout, we use ``$\log$'' to denote the natural logarithm. Whenever the wavenumber, $k$, appears in a mathematical expression, it is assumed to have units of $\hMpc$.

\section{Symbolic Emulators for \texorpdfstring{$\Lambda$}{Lambda}CDM}
\label{sec:wider syren}

The aim of this section is to extend some of the SYmbolic Regression ENhanced (\syren) \cite{Bartlett_2024_linear,Bartlett_2024_syren,Sui_2024} emulators so that they are valid over a wider parameter range than in the original papers.
For simplicity, we will focus on the concordance cosmological model ($\Lambda$CDM) which is governed by six key parameters: the present day baryon and total matter densities ($\Omega_{\rm b}$ and $\Omega_{\rm m}$, respectively), the Hubble constant  ($H_0 = 100 h {\rm \ km/s/Mpc}$), the reionisation optical depth, $\tau$, and two parameters describing the amplitude, $A_{\rm s}$, and spectral tilt, $n_{\rm s}$, of the primordial power spectrum.
In this work we ignore the small effect of $\tau$ on the quantities of interest and often use an alternative parametrisation where we replace $A_{\rm s}$ with $\sigma_8$, which is the root mean squared fluctuation of the linearly evolved density field when smoothed with a top-hat filter of radius $8 \,\Mpch$ (although we provide an emulator for converting between the two parametrisations).
For these parameters, we use the prior range given in \cref{tab:parameter_prior}, since this is more appropriate for current cosmological analyses.
Given that we are particularly focussing on weak-lensing analyses, we fit our models for redshifts $z \leq 3$, although we demonstrate below that our models have excellent extrapolation behaviour beyond this range. 
We also produce some new emulators for quantities which are required for a weak lensing analysis and whose evaluation can be slow, 
namely, the comoving distance and linear growth factor.

In the remainder of this section, we begin by describing the symbolic regression code used (\cref{sec:operon}) before outlining the quantities we wish to emulate and our found symbolic approximations (\cref{sec:emulators}). 
We assess the performance of our emulators at the level of the nonlinear matter power spectrum in \cref{sec:halofit_test}.
We note that, due to factorising out their dependence, all the emulators developed in this section can be applied to an arbitrary $\sigma_8$ or $A_{\rm s}$ except for the \halofit{} ones, provided the other parameters do not go outside of the ranges given in \cref{tab:parameter_prior}. 

\begin{table}
    \centering
    \begin{tabular}{c|c|c|c}
         Parameter &  Description  & Min Value & Max Value \\
         \hline
         \hline
        $\Omega_{\rm m}$  & Present-day matter density parameter & 0.1 & 0.5 \\
        $\Omega_{\rm b}$ & Present-day baryon density parameter & 0.03 & 0.07 \\
        $h$ & Hubble constant ($H_0 = 100 h {\rm \ km/s/Mpc}$) & 0.5 & 0.9 \\
        $n_{\rm s}$ & Scalar spectral index & 0.8 & 1.2 \\
        $\sigma_8$ & Matter fluctuation amplitude  & 0.6 & 1.0 \\
        $z$ & Redshift & 0.0 & 3.0 \\
        $k$ & Wavenumber (\hMpc) & $10^{-4}$ & $10^2$
    \end{tabular}
    \caption{Prior range used to generate symbolic emulators in this work.}
    \label{tab:parameter_prior}
\end{table}

\subsection{Symbolic Regression}
\label{sec:operon}

Our symbolic emulators are obtained using the genetic programming-based symbolic regression code \operon{} \cite{Burlacu_2020}. 
Genetic programming is an evolutionary algorithm that operates on symbolic expressions represented as trees. In each generation, a population of candidate expressions is evaluated using a fitness function, and less effective expressions are removed. New candidates are generated through crossover -- recombining parts of existing expression trees -- and mutation, which involves randomly altering tree structures by inserting, deleting, or replacing subtrees. Through successive generations, the expression population is expected to improve, yielding increasingly accurate analytic models.
In \operon, the leaf nodes of expressions (a variable or a parameter) are paired with associated scaling factors. These scaling coefficients are subsequently optimised using the Levenberg–Marquardt algorithm \cite{Levenberg_1944,Marquardt_1963}, as described in \cite{Kommenda_2020}.

Our Pareto-optimisation problem attempts to find an accurate fitting function while remaining relatively simple. During the search we measure simplicity as the model `length', which is equal to the number of nodes in the tree-representation of the function, and accuracy with the root mean squared error (RMSE). 
Unless otherwise specified, we seek equations which can be formed from the following symbols: $+$, $-$, $\times$, $\div$, $\sqrt{\cdot}$, ${\rm pow}$ and $\log$, as well as constants and variables.
\operon{} incorporates $\epsilon$-dominance during the non-dominated sorting stage \cite{Laumanns_2002}, where two candidate solutions are 
considered equivalent if their objective values—accuracy and simplicity—differ 
by less than a specified threshold $\epsilon$. This approach limits the number 
of near-identical individuals in the population and supports convergence 
towards a diverse approximation of the Pareto front: the set of solutions for 
which no improvement in accuracy is possible without increasing complexity.

To balance predictive performance with model compactness, we tested several 
$\epsilon$ values and report the selected settings in the relevant sections for 
each emulator. Genetic programming based symbolic regression is a nondeterministic method, and thus identical setups but with different initial populations can lead to very different expressions and loss functions. As such, the variation in our results with different values of $\epsilon$ could also be due to different initial populations. Nonetheless, our final chosen settings and initial populations yield sufficiently accurate expressions for our purpose, although the reader could obtain different expressions (of better or worse accuracy) if they reran our analysis with a different initial random seed, even if they used the same training data.

To select our models, we manually inspect all equations returned on the Pareto front after a specified run time. We reject all equations which we deem to be insufficiently accurate for our purpose or whose losses on the training and validation sets are significantly different; an indication of over-fitting. The selection of our preferred model is then somewhat qualitative based on the overall structure of the model (although note that this can be quantified using language models \cite{Bartlett_2023}), but we also only select models which have (or can be forced to have) the correct analytic limits in known cases.

\subsection{Emulators}
\label{sec:emulators}

\subsubsection{Radial comoving distance}
\label{subsec:comoving_distance}

In $\Lambda$CDM, the radial comoving distance to an object at scale factor $a \equiv 1 / (1 + z)$ is
\begin{equation}
    \label{eq:chi_exact}
    \chi(a) = \frac{c}{H_0} \int_a^1 \frac{{\rm d} a^\prime}{{a^\prime}^2 \sqrt{\Omega_{\rm m} {a^\prime}^{-3} + 1 - \Omega_{\rm m}}}
    = \frac{c}{H_0}\left[ \frac{2 a^\prime}{\Omega_{\rm m}} \hyper \left(\frac{2}{3}, 1, \frac{7}{6}; x^\prime \right) \sqrt{\Omega_{\rm m} {a^\prime}^{-3} + 1 - \Omega_{\rm m}}  \right]_{a^\prime = a}^1,
\end{equation}
where $x^\prime \equiv {a^\prime}^3 (\Omega_{\rm m} - 1) / \Omega_{\rm m}$ and $\hyper{}$ is the Gaussian hypergeometric function.
We can interpret $-x^\prime$ as the ratio of the energy density of dark energy to that of matter at the scale factor $a^\prime$.
Although this has a closed-form solution, this is not always practical for modern computational workflows. In particular, libraries such as \textsc{jax.scipy} (at the time of writing) do not provide built-in support for the hypergeometric function in \cref{eq:chi_exact}, and its direct evaluation can be computationally inefficient. This presents a challenge for applications that rely on automatic differentiation, such as those using \textsc{jax}, where efficient and differentiable expressions are preferred.
We therefore wish to find a simpler analytic approximation for $\chi(a)$.

Given \cref{eq:chi_exact}, we only need to find an approximation for the hypergeometric function part, since the other terms are trivial to evaluate.
To obtain an approximation for the prior range of interest (\cref{tab:parameter_prior}), we need to have an approximation which is valid for the range $-9.0 \leq x^\prime \leq -0.016$.
We therefore generated 200 $x$ points randomly from a uniform distribution in this range for training, and another 200 for validation, and evaluate the hypergeometric function using \textsc{scipy}.
We search for expressions with a maximum length of 25 and terminate \operon's search after 2 minutes on 40 cores. We choose $\epsilon=10^{-5}$ and add the analytic quotient operator ($\aq (x, y) \equiv x / \sqrt{1 + y^2}$) to the default basis set outlined in \cref{sec:operon}.

The longest model which \operon{} returns with these settings is at length 15, where our approximation to $\hyper ( \frac{2}{3}, 1, \frac{7}{6}; x )$ is proportional to $(c_0 + \left(c_{1} - x\right)^{c_{2}} - c_3 x)^{- c_{4}}$, where $\{c_i\}$ are optimised parameters. 
Given its relative simplicity and that it is the most accurate model found, we choose this expression.
\operon{} returns a constant of proportionality, however we choose not to use this value but instead explicitly enforce that $\hyper ( \frac{2}{3}, 1, \frac{7}{6}; 0) = 1$. This gives
\begin{equation}
    \label{eq:chi_approx}
    \hyper \left(\frac{2}{3}, 1, \frac{7}{6}; x \right) \approx \left(\frac{c_{0} + c_{1}^{c_{2}}}{ c_0 + \left(c_{1} - x\right)^{c_{2}} - c_3 x} \right)^{c_{4}},
\end{equation}
for $\{c_i\} = \splitatcommas{\{0.9207, 0.98617, 1.42499, 0.91875, 0.46516\}}$.

\begin{figure}
    \centering
    \includegraphics[width=\textwidth]{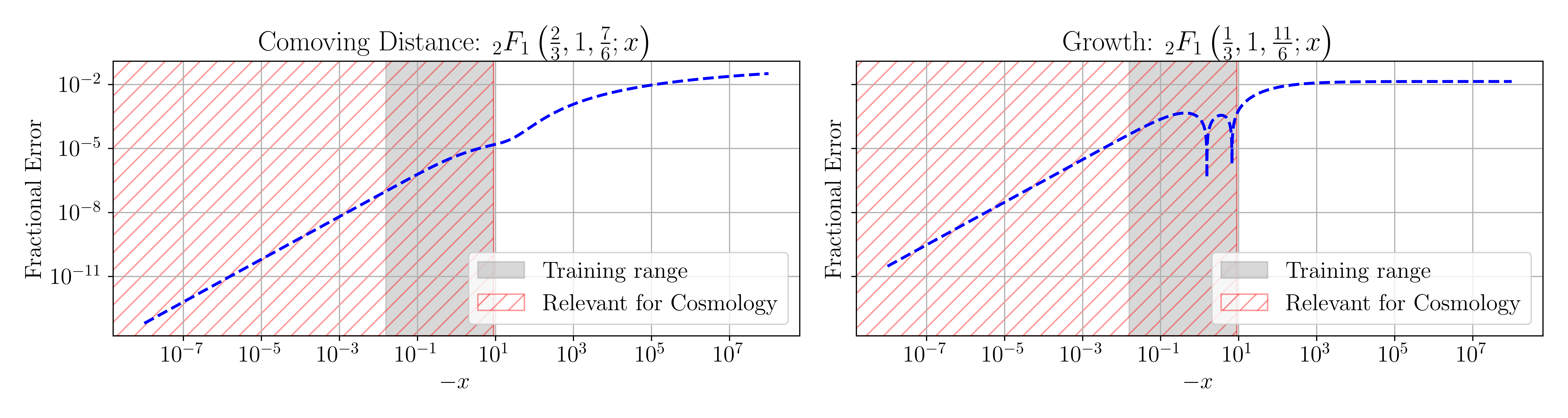}
    \caption{Fractional errors on our approximations to the hypergeometric functions required to evaluate the radial comoving distance (left, \cref{eq:chi_approx}) and linear growth factor (right, \cref{eq:D_approx_lcdm}) for a $\Lambda$CDM cosmology. 
    In both cases, $x \equiv a^3 (\Omega_{\rm m} - 1) / \Omega_{\rm m}$, for scale factor $a$ and present-day matter density parameter $\Omega_{\rm m}$.
    The approximations were obtained for $x$ in the grey shaded region (corresponding to the prior range in \cref{tab:parameter_prior}), and they thus show very good extrapolation behaviour, particularly for values of $x$ relevant for cosmology (red hatched region).
    }
    \label{fig:hyper_error}
\end{figure}

We see that this is an exceptionally good approximation in \cref{fig:hyper_error}, where we plot the fractional error on this approximation as a function of $-x$. When considering the range of the training data, the fractional error monotonically increases with $|x|$, from approximately $10^{-7}$ for small $|x|$ to $10^{-5}$ at the largest value of $x$, and thus the error is negligible.
In \cref{fig:hyper_error} we also show the extrapolation behaviour of this function.
Given that we only consider positive redshifts in cosmology, we are only concerned with the extrapolation to smaller values of $|x|$. Remarkably, the error continues to decrease as we consider higher redshifts, and thus for all $0.1 < \Omega_{\rm m} < 0.5$ and for any redshift, we always have errors smaller than 0.001\% with this approximation. The error slightly increases if we consider the other (irrelevant) limit for extrapolation, but still remains sub-percent until we consider values of $x$ which are $10^4$ times larger than the maximum value in our training range.

\subsubsection{Growth factor}

We now turn to the linear growth factor. This function gives the time dependence of perturbations at linear order in cosmology, and is thus of fundamental importance.
The growth factor in a $\Lambda$CDM universe is given by \cite{Heath_1977,Peebles_1980}
\begin{equation}
    D(z) =  \hyper \left(\frac{1}{3}, 1, \frac{11}{6};  x\right) a.
\end{equation}
where $x \equiv a^3 ( \Omega_{\rm m} - 1) / \Omega_{\rm m}$.
We search for an analytic approximation to this, again using 200 training and validation points in the range used for the emulator of $\chi(a)$ and with the same basis functions as in the preceding section. After only 10 seconds of searching, \operon{} finds a model at length 11
which is proportional to $( \left(b_{0} - x\right)^{b_{2}} + b_1)^{-1/2}$, for some constants $\{b_i\}$. Although \operon{} gives the constant of proportionality, we choose explicitly enforce that $\hyper \left(\frac{1}{3}, 1, \frac{11}{6}; 0 \right) = 1$, which corresponds to ensuring $D(a) \approx a$ at early times. We also know that, for large but negative $|x|$, the hypergeometric function is proportional to $(-x)^{-1/3}$, which sets the value of $b_2$ to be $2/3$.

After making these adjustments, we re-optimise the parameters by minimising the RMSE across our training set, yielding the final expression
\begin{equation}
    \label{eq:D_approx_lcdm}
    \hyper \left(\frac{1}{3}, 1, \frac{11}{6}; x \right) \approx \frac{\sqrt{b_{0}^{2/3} + b_1}}{\sqrt{\left(b_{0} - x\right)^{2/3} + b_1}},
\end{equation}
where $\{b_i\} = \{ \splitatcommas{0.723, 1.204} \}$.

We plot the fractional error on this approximation as a function of $|x|$ in \cref{fig:hyper_error}, where we again see excellent results. For the range of the training data (i.e. for the range of parameters in \cref{tab:parameter_prior}), the fractional error on this expression 
is always below $5 \times 10^{-4}$,
so is once again negligible. The extrapolation behaviour is the same as before: this approximation improves as one extrapolates to higher redshift and thus the error is negligible for all values of $x$ relevant to cosmology. For the other limit, the error increases, but stays at approximately percent level even for much larger values of $x$ than \cref{eq:chi_approx}.

\subsubsection{Conversion between \texorpdfstring{$A_{\rm s}$}{As} and \texorpdfstring{$\sigma_8$}{sigma8}}
\label{sec:As emulator}

Our remaining emulators are for functions and variables which describe the clustering of matter in the Universe. The matter density of the Universe, $\rho(\bm{x})$, can be factorised into a spatially constant background density, $\bar{\rho}$, and a density contrast $\delta ( \bm{x})$, such that $\rho(\bm{x}) \equiv \bar{\rho} ( 1 + \delta(\bm{x}))$.
We then define the Fourier transform of $\delta ( \bm{x})$ to be $\tilde{\delta}(\bm{k})$ and assume that the distribution of matter is statistically homogeneous and isotropic. Defining $\langle\cdots\rangle$ to be an ensemble average, the matter power spectrum, $P(k)$, is defined to be
\begin{equation}
    (2 \pi)^3 P(k,\bm{\theta}) \delta^{\rm D} \left( \bm{k} - \bm{k}^\prime \right)\equiv\langle \tilde{\delta}(\bm{k}) \tilde{\delta}^\ast (\bm{k}^\prime) \rangle,
\end{equation}
where $\delta^{\rm D}$ is the Dirac delta function. If the density constrast is predicted according to linear perturbation theory, then we call this the linear matter power spectrum, $P_{\rm lin}(k, \bm{\theta})$.

An important quantity is the root mean squared fluctuations of these linearly evolved fluctuations within spheres of radius $R$. This is given by
\begin{equation}
    \label{eq:sigmaR}
    \sigma_R^2 (z, \bm{\theta}) = \int_0^\infty {\rm d}k \, \frac{k^2}{2 \pi^2} P_{\rm lin} (k, z, \bm{\theta}) \left| W(k,R) \right|^2,
\end{equation}
where $W(k,R)$ is the Fourier transform of the top-hat window function and is given by
\begin{equation}
    W(k, R) = \frac{3}{(kR)^3} \left( \sin (k R) - kR \cos (k R) \right).
\end{equation}
It is common in cosmology to replace the parameter $A_{\rm s}$ by $\sigma_8$, which is simply $\sigma_R$ for $R = 8 \Mpch$ at $z=0$. However, if we want to change from one parametrisation to another, it is convenient to have a simple parametric expression for this conversion rather than having to evaluate and integrate the linear power spectrum.

Since the linear power spectrum is proportional to $A_{\rm s}$, we know that $\sqrt{A_{\rm s}}/\sigma_8$ can only depend on the other cosmological parameters: $\splitatcommas{\Omega_{\rm m}, \Omega_{\rm b}, h, n_{\rm s}}$. We therefore with find an approximation for $\log \left(\sqrt{10^9 A_{\rm s}}/\sigma_8 \right)$ as a function of these parameters, where we multiply $A_{\rm s}$ by $10^9$ to obtain a target of order 1, and take the logarithm so that we always predict a positive $\sigma_8$ for a given $A_{\rm s}$.
To find this function, we generate 200 training cosmologies using a Latin hypercube and the parameter prior in \cref{tab:parameter_prior}. For a given $A_{\rm s}$, we then find the corresponding $\sigma_8$ using \camb{} \cite{Lewis_2000} and compute $\log \left(\sqrt{10^9 A_{\rm s}}/\sigma_8 \right)$. We repeat this process for a further 200 cosmologies from a different Latin hypercube, which we use as our validation set. We search for expressions with \operon{} using the default function set from \cref{sec:operon} with $\epsilon=10^{-6}$ and with a maximum allowed model length of 50.

After 5 minutes of searching on a single node of 40 cores, we find that the loss for the Pareto front plateaus at a model length of 42, and thus choose this model, which is
\begin{equation}
    \label{eq:As_fit}
    \begin{split}
        \log \left( \frac{\sqrt{10^9 A_{\rm s}}}{\sigma_8}\right) &\approx 
    d_{0} n_{s} \left(d_1 \Omega_{\rm m}\right)^{- d_{2} h} 
    + \left(d_{3} \sqrt{h} + d_{4} n_{\rm s}\right) 
    \left(- d_{5} n_{s} + \frac{\Omega_{\rm b} - d_{6}}{d_7 \Omega_{\rm m} - d_8 \Omega_{\rm b}}\right)  \\
    &  + \left(d_{9} \Omega_{\rm m}\right)^{- d_{10} h}+ \frac{\left(d_{11}\Omega_{\rm m}\right)^{d_{12} h} \left(d_{13}\Omega_{\rm b} + \left(d_{14} h\right)^{- d_{15}\Omega_{\rm m}}\right)}{d_{16} \Omega_{\rm m} + d_{17} \Omega_{\rm b} },
    \end{split}
\end{equation}
where the parameters $\{d_i\}$ are given in \cref{tab:As_par}.
This model is very accurate, with a RMSE of $1.3 \times 10^{-3}$ and $1.6 \times 10^{-3}$ on the training and validation sets, respectively. Given that we often seek percent-level accurate predictions for the power spectrum, this is far within the required accuracy, since the fractional error on the power spectrum is approximately twice the error on the fit in \cref{eq:As_fit}
(since we have $\sqrt{A_{\rm s}}$ on the left-hand side of \cref{eq:As_fit}). This error is comparable to the result of \cite{Bartlett_2024_linear} despite the larger prior range considered here.

\begin{table}
  \centering
  \begin{tabular}{c c | c c | c c | c c | c c}
 Name & Value & Name & Value & Name & Value & Name & Value & Name & Value \\
    \hline\hline
    $d_{0}$ & 0.95534 & $d_{1}$ & 68.8078 & $d_{2}$ & 0.5159 & $d_{3}$ & 1.18861 & $d_{4}$ & 0.197 \\
    $d_{5}$ & 0.53884 & $d_{6}$ & 0.01983 & $d_{7}$ & 0.76405 & $d_{8}$ & 0.29247 & $d_{9}$ & 10.8834 \\
    $d_{10}$ & 0.73004 & $d_{11}$ & 1.20497 & $d_{12}$ & 0.75788 & $d_{13}$ & 2.14175 & $d_{14}$ & 3.03762 \\
    $d_{15}$ & 4.71485 & $d_{16}$ & 5.46729 & $d_{17}$ & 0.9624 \\
  \end{tabular}
  \caption{Best-fit parameters for the emulator converting $A_{\rm s}$ to $\sigma_8$ (\cref{eq:As_fit}).}
  \label{tab:As_par}
\end{table}

\subsubsection{Linear matter power spectrum}

One of the most costly parts of the analysis that we will perform in \cref{sec:cosmo_analysis} is evaluating the linear matter power spectrum, and thus an emulator for this is the most impactful for speeding up the analysis.
This acceleration is not limited to weak lensing analyses; for example, an Effective Field Theory based analysis of galaxy clustering also requires the linear matter power spectrum, so this could be easily incorporated into such an analysis.

As in previous work \cite{Bartlett_2022,Sui_2024}, to obtain an approximation for this, we begin with the approximation from \eh{} \cite{Eisenstein_1998} without the baryonic acoustic oscillations (BAOs) ($P_{\rm no \, wiggle}(k, \bm{\theta})$), and learn the ratio of the true linear power spectrum (evaluated using \camb) to this
\begin{equation}
    P_{\rm lin}(k, z, \bm{\theta}) = D^2(z, \bm{\theta}) P_{\rm no \, wiggle} (k, \bm{\theta}) F (k, \bm{\theta}).
\end{equation}
As such, $F(k, \bm{\theta})$ fully captures the BAOs, but also provides a correction to the no-wiggle part of the \eh{} prediction.
We choose to learn $\log F$ to enable us to optimise the fractional error and to ensure that our result for $F$ is always positive. Since we are dealing with $\Lambda$CDM, $F$ is independent of redshift \cite{Dodelson_2003}. We also know that it is independent of $A_{\rm s}$ since its dependence is fully captured by $P_{\rm no \, wiggle}$, and thus only fit this ratio as a function of $\Omega_{\rm m}$, $\Omega_{\rm b}$, $h$, $n_{\rm s}$ and $k$.

To perform this fit, we sample 200 cosmologies using a Latin hypercube in the range given by \cref{tab:parameter_prior}. We use a further 200 for validation. For each cosmology, we evaluate the matter power spectrum at 200 $k$ values, although these are not evenly spaced in $k$ nor $\log k$.
We wish to have high resolution near the oscillatory features of the power spectrum (BAOs), and thus use 150 values of $k$ logarithmically spaced between $10^{-2} \, h \, {\rm Mpc}^{-1}$ and $1 \, h \, {\rm Mpc}^{-1}$. The remaining 50 are used in the rest of the $k$ range specified in \cref{tab:parameter_prior} such that the logarithmic spacing is constant for those two regions. Outside the high-resolution region, $\log F$ varies slowly, hence we are able to use fewer points.

This is our most challenging emulator to fit, and thus we choose a significantly longer run time for \operon{} of 47 hours on 128 cores (after 24 hours our results were slightly too inaccurate for our purposes). For numerical stability, we replace division in our function set with $\aq$ (again defined to be $\aq (x, y) \equiv x / \sqrt{1 + y^2}$), and we also include cosine since we know that we will have to fit oscillatory features. We set  $\epsilon=10^{-4}$ and allow a maximum model length of 200.

We find that \operon{} produces a particularly compact result at a model length of 167, which takes the form
\begin{equation}
    \label{eq:pk_lin_fit}
    \log F (k, \bm{\theta} )\approx f_1 \left( f_2 \cos(f_3) + f_4\right) + f_5,
\end{equation}
where each $\{f_i\}$ are functions of $k$ and $\bm{\theta}$, except $f_5$ which only depends on $\bm{\theta}$. These are given in \cref{app:pk_lin}.
We note that the form of each of $\{f_i\}$ are non-trivial functions of $k$ and the cosmological parameters. As such, although we are able to interpret \cref{eq:pk_lin_fit} in various limits below, this is not generally true for arbitrary $k$ from just looking at the equations. Nonetheless, this fit is accurate and, once the results are plotted, we can begin to see how each of the terms in \cref{eq:pk_lin_fit} correct the \eh{{} formula.}

\operon{} returns a constant for $f_5$, however we choose not to use this since we must obey a physical constraint. Since the \eh{} transfer function is unity on large scales, we require that $\log F \to 0$ as $k\to 0$. We therefore analytically compute $\tilde{f}_i \equiv \lim_{k \to 0} f_i$ and, to obey the physical constraint, enforce that
\begin{equation}
    f_5 = - \tilde{f}_1 \left( \tilde{f}_2\cos(\tilde{f}_3) + \tilde{f}_4\right).
\end{equation}
It is not guaranteed \textit{a priori} nor forced in the search that the limits of these expressions as $k \to 0$ are well behaved, however for our chosen model each $f_i$ is finite in this limit.
We find that $\tilde{f}_2=1$ whereas the other $\tilde{f}_i$ depend on the cosmological parameters.

\begin{figure}
    \centering
    \includegraphics[width=\textwidth]{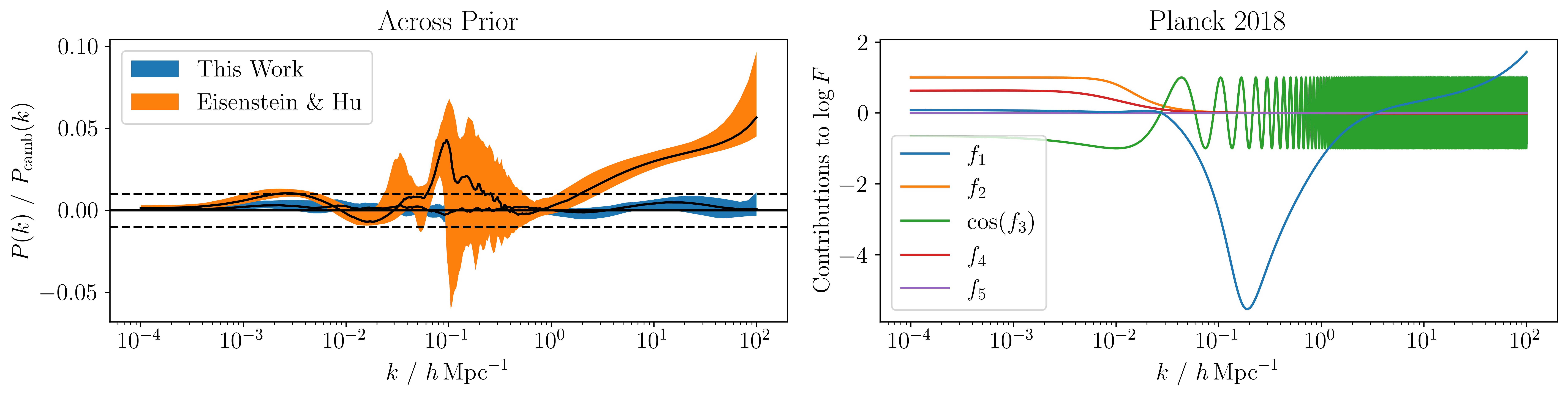}
    \caption{\textit{Left}: Fractional error on symbolic fits to the redshift-zero linear matter power spectrum when compared against \camb. We plot the 68\% error distributions when the cosmological parameters are varied across the range given in \cref{tab:parameter_prior}. \textit{Right}: The various contributions to the prediction for $\log F$ at the Planck 2018 cosmology. The high frequency oscillations of $\cos(f_3)$ are a plotting artefact from using a logarithmic $x$ scale: for large $k$, $f_3$ is linear in $k$.
    }
    \label{fig:pk_lin}
\end{figure}

We compute the error on this fit compared to \camb{} as a function of $k$ and averaged across the prior range of \cref{tab:parameter_prior}, and plot the results in \cref{fig:pk_lin}. We compare the results to the errors on the \eh{} formula which includes the wiggles \cite{Eisenstein_1999} as implented in \jaxcosmo{} \cite{Campagne_2023}. For both cases, we only use the transfer function with the true $A_{\rm s}$ rather than integrating and changing the normalisation so that $\sigma_8$ matches.
One immediately sees that the approximation obtained here is much more accurate than that of \eh; when averaged across all cosmologies and wavenumbers of \cref{tab:parameter_prior}, we obtain a RMSE of 0.6\% error, compared to 3\% for \eh. Moreover, from \cref{fig:pk_lin} we see that the error on our fit is relatively constant with wavenumber, but is a strong function of $k$ for \eh; our approximation's error band stays within 1\% for all $k$, whereas the \eh{} result grows to up to 10\% error on the smallest scales.

In \cref{fig:pk_lin}, we also illustrate the various components contributing to the symbolic fit, using the Planck 2018 cosmological parameters. Each term in the model has a distinct interpretative role. The function $f_1$ accounts for the high-$k$ inaccuracies of the \eh{} fitting formula, which arise due to it neglecting baryon pressure effects \cite{Eisenstein_1998} on small scales. The terms $f_2$ and $f_4$ act as approximate step functions that effectively localise the oscillatory features to a finite range in $k$, suppressing them outside this window. The oscillatory component $\cos(f_3)$ behaves approximately as $\cos(\omega k + \phi)$, where $\omega$ and $\phi$ are asymptotically constant in the small- and large-$k$ limits, with smooth corrections in the intermediate regime. Finally, $f_5$ enters the expression as a constant offset to ensure the correct limit as $k \to 0$.

\subsubsection{Halofit variables}

Up until now, we have only considered the linear evolution of matter in the Universe. Although a good approximation on large scales, for smaller scales one cannot neglect the nonlinear effects of gravity, and thus one must correct the matter power spectrum. This has the effect of distorting the BAOs and introducing a non-trivial time dependence, such that $P(k)$ is larger on small scales and the ratio of the nonlinear to linear power spectrum on these scales grows with time.

The most accurate way to capture this behaviour is through running $N$-body simulations, however these are far too computationally expensive to be used within an inference pipeline, and thus surrogate models are required to estimate the effects of nonlinearity on $P(k)$. Many numerical emulators based on these have been created \cite{Angulo_2021,Zennaro_2023,Knabenhans_2019,Knabenhans_2021}, and the \textsc{syren-halofit} \cite{Bartlett_2024_syren} and \textsc{syren-new} \cite{Sui_2024} symbolic emulators have been shown to demonstrate similar accuracy to these methods but at a fraction of the computational cost per evaluation. Unfortunately, these emulators were trained on $N$-body simulations with prior ranges smaller than that of \cref{tab:parameter_prior}, and thus in this work we will not consider these emulators further.

Instead, we wish to emulate a semi-analytic prescription of nonlinear physics, namely \textsc{halofit} \cite{Smith_2003,Bird_2012,Takahashi_2012}. 
In this work we use the commonly-used parameters of \cite{Takahashi_2012}, which are optimised to match the results of 16 $N$-body simulations around the Wilkinson Microwave Anisotropy Probe (WMAP) cosmologies \cite{Spergel_2003}, although we note an updated set of parameters are presented in \cite{Bartlett_2024_syren}.

\halofit{} gives the nonlinear matter power spectrum as a function of the linear spectrum, the cosmological parameters, as well as three derived variables computed from the linear spectrum. The first of the these is the nonlinear scale, $k_\sigma$, which is defined to be
the wavenumber
at which
\begin{equation}
    \label{eq:ksigma definition}
    \sigma_{\rm G}^2(k_\sigma^{-1}) \equiv 1, \quad
    \sigma_{\rm G}^2(R) \equiv \int \Delta_{\rm lin}^2 (k) \exp\left(-k^2 R^2\right) \, {\rm d} \log k,
\end{equation}
for
\begin{equation}
    \Delta_{\rm lin}^2 (k) \equiv P_{\rm lin}(k) \frac{k^3}{2 \pi^2},
\end{equation}
and $P_{\rm lin}(k)$ is the linear matter power spectrum at the given redshift.
Once we have $k_\sigma$, we also define
\begin{equation}
    \label{eq:neff C definition}
    n_{\rm eff} + 3 \equiv - \left. \frac{{\rm d} \log \sigma_{\rm G}^2(R)}{{\rm d} \log R} \right|_{\sigma_{\rm G} = 1},   \quad
    C \equiv - \left. \frac{{\rm d}^2 \log \sigma_{\rm G}^2(R)}{{\rm d} \log R^2} \right|_{\sigma_{\rm G} = 1},  
\end{equation}
where the derivatives can be computed by fitting splines to $\sigma_{\rm G}^2(R)$.

Obtaining these variables can be slow to evaluate since one needs to perform integrals of the linear power spectrum, run a root-finding algorithm, fit splines and then take derivatives of these. As such, the \textsc{syren-halofit} emulators \cite{Bartlett_2024_syren} find simple analytic expressions for $k_{\sigma}$, $n_{\rm eff}$ and $C$ as a function of cosmological parameters and redshift. In this section we extend these emulators to the prior of \cref{tab:parameter_prior}. Given the wider prior range than \textsc{syren-halofit}, we require more training samples than in that work; we use 1000 for each variable rather than 200, with parameters sampled uniformly on a LH given the prior range of \cref{tab:parameter_prior}.
We compute $k_\sigma$ by running a root-finding algorithm to find the value of $R$ which obeys \cref{eq:ksigma definition}, where we compute the linear power spectrum with \camb. Once this is obtained, we find $\sigma_{\rm G}^2(R)$ with a fifth order smoothing spline, and take derivatives of this to compute $n_{\rm eff}$ and $C$.
To approximate these variables, we seek expressions with a maximum length of 100 and comprised of the symbols $\splitatcommas{+,-,\times, \div, \sqrt{\cdot}, {\rm pow}, \log}$ as well as constants and variables.

\begin{figure}
    \centering
    \includegraphics[width=\textwidth]{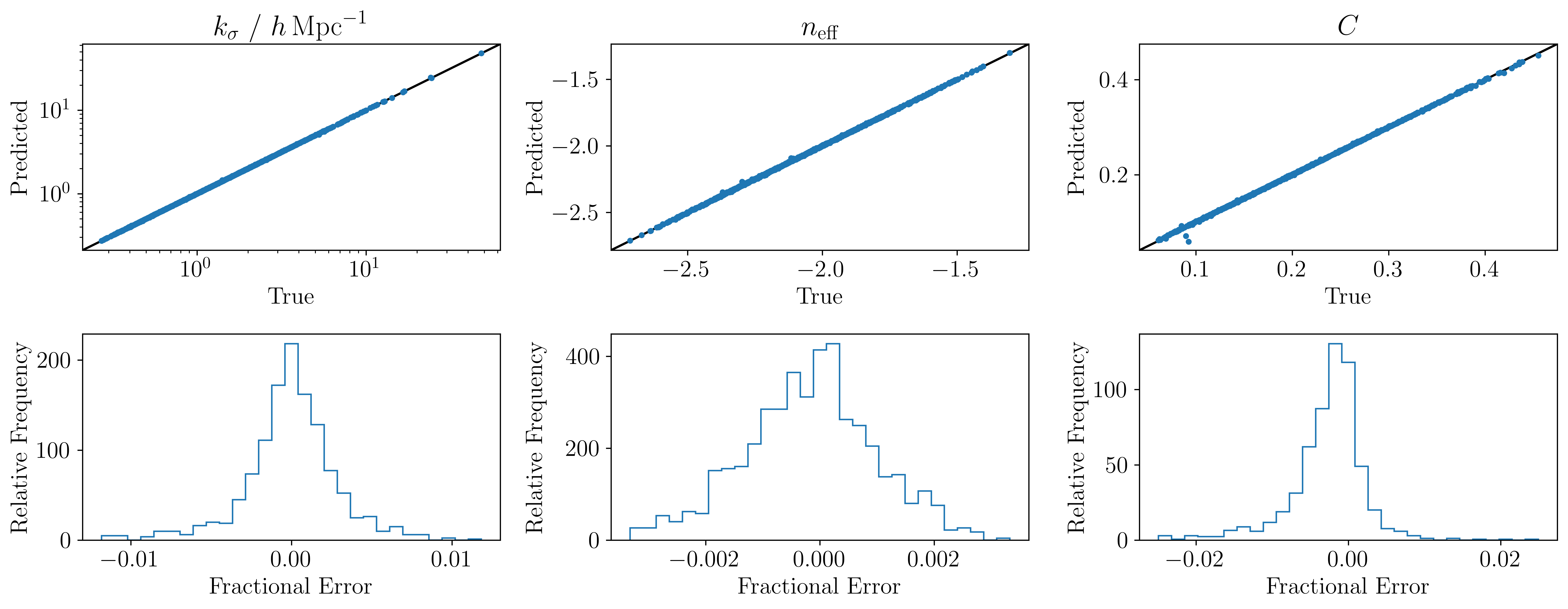}
    \caption{Comparison between the true and symbolic fits for the halofit variables: the nonlinear scale ($k_\sigma$; left), the effective slope ($n_{\rm eff}$; centre), and curvature ($C$; right). We plot the predicted vs true values in the upper panels and the distribution of their fractional errors in the lower panels.
    }
    \label{fig:test_halofit_vars}
\end{figure}

Beginning with $k_\sigma$, we find that, given the large dynamical range of values, it is preferable to fit $\log k_\sigma$ rather than $k_\sigma$. This has the additional benefit that $k_\sigma$ is guaranteed to be positive. We optimise jointly the model length and mean absolute error (MAE) on  $\log k_\sigma$ (corresponding to a fractional error on $k_\sigma$), and choose $\epsilon=10^{-5}$. After searching for 24 hours with \operon{}  using a single node with 128 cores, we choose the model at length 95, which is given in \cref{eq:ksigma_fit}.
We plot the distribution of fractional errors on this fit in \cref{fig:test_halofit_vars}. We find that this has mean absolute values of 
$2.0 \times 10^{-3}$ and $2.3 \times 10^{-3}$ for the training and validation sets, respectively, and is thus accurate to much greater than percent-level accuracy.

We find that $n_{\rm eff}$ is the easiest of the \halofit{} variables to fit, requiring a model of only length 44. This model was found after 24 hours of searching with \operon{} on 128 cores, again with $\epsilon = 10^{-5}$. However, on this occasion we fit for $n_{\rm eff}$ directly and choose to optimise the RMSE instead of MAE. This model is given in \cref{eq:neff_fit} and the distribution of errors plotted in \cref{fig:test_halofit_vars}.
This model is not only the shortest of the \halofit{} variables, but is also the most accurate, with mean absolute fractional errors of $8.9 \times 10^{-4}$ and $9.9 \times 10^{-4}$ over the training and validation sets, respectively.

For our final \halofit{} variable, $C$, we find that we only need 12 hours of run time with the same resources to find an appropriate model with \operon, which was found by jointly minimising the MAE on $C$ and the model length. We again choose $\epsilon=10^{-5}$. The chosen model has a length of 71 and is given in \cref{eq:C_fit}.
 We find that this is the least accurate of the three \halofit{} variables, with mean absolute fractional errors on $4.9 \times 10^{-3}$ and $4.7 \times 10^{-3}$ on the training and validation sets, respectively. Nonetheless, this is still within percent level accuracy for the vast majority of the training and validation set, as can be seen in \cref{fig:test_halofit_vars}.

\subsection{Nonlinear matter power spectrum prediction}
\label{sec:halofit_test}

\begin{figure}
    \centering
    \includegraphics[width=\textwidth]{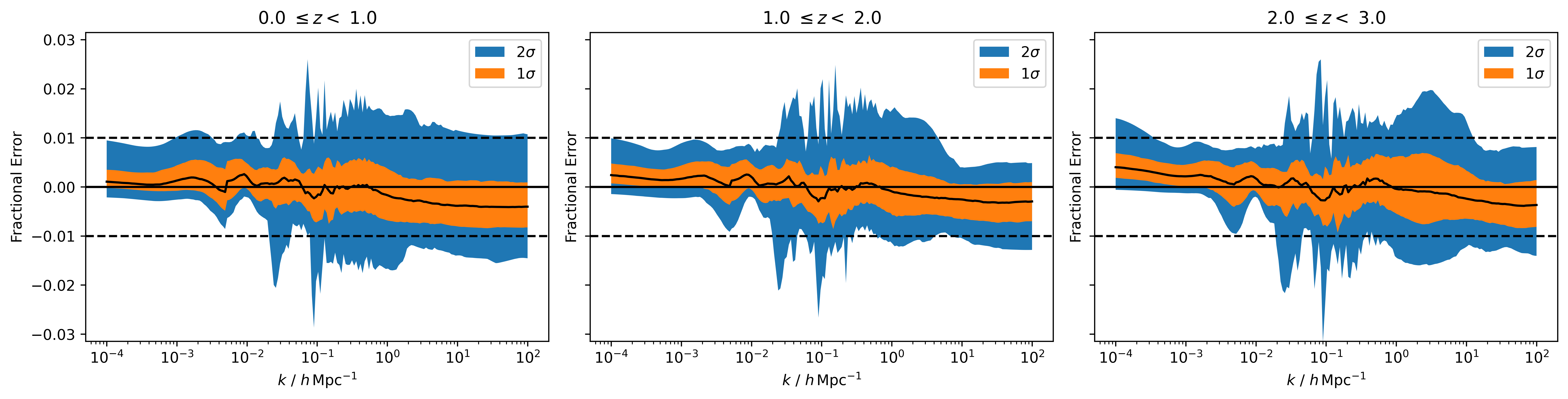}
    \caption{Fractional error on symbolic fits to the \halofit{} approximation to the nonlinear matter power spectrum when compared against \camb's implementation of \halofit. We plot the 68\% and 95\% error distributions when the cosmological parameters are varied across the range given in \cref{tab:parameter_prior}, and in each panel we choose randomly sampled redshifts in the ranges given by the titles. Our approximation utilises the symbolic approximation for the linear matter power spectrum (\cref{eq:pk_lin_fit}), the growth factor (\cref{eq:D_approx_lcdm}) and the \halofit{} variables (\cref{eq:ksigma_fit,eq:neff_fit,eq:C_fit}).
    }
    \label{fig:test_halofit_spectra}
\end{figure}

Given the nontrivial dependence of the final \halofit{} variables on the prediction for the nonlinear matter power spectrum, we now investigate how these emulators perform when predicting the full $P(k)$. 
We consider the Takahashi et al. 2012 \cite{Takahashi_2012} implementation of \halofit, and compare the results of using our approximations for the growth factor, linear power spectrum and \halofit{} variables within this framework to that obtained using \camb.

Utilising a Latin hypercube of parameters and redshifts in the range given in \cref{tab:parameter_prior}, in \cref{fig:test_halofit_spectra} we plot the distributions of fractional errors on the \halofit{} $P(k)$ as a function of $k$, and split into three equal redshift ranges.
We see that the errors on our predictions are relatively independent of $z$, with all three panels of \cref{fig:test_halofit_spectra} appearing very similar. We observe that the 68\% error band is always within 1\% for all $k$ and $z$, and that the 95\% band is often at or within this level. As such, we see that our emulators are able to produce percent-level accurate predictions for this quantity, even when combined in this way.

\section{Mock cosmological analysis}
\label{sec:cosmo_analysis}

Until now, we have focused on the performance of the individual symbolic emulators at the level of the quantities which they are emulating. Although we have seen good performance, it is essential to assess how well they perform in a cosmological inference; even if the accuracy levels appeared high in previous sections, if they lead to biased cosmological inferences then they would be unusable.
In this section we perform this test, where we perform a mock analysis of a Dark Energy Survey Year 1 (DES-Y1) lensing and clustering cosmological analysis \cite{Abbott_2018,Krause_2017}. We describe the setup of our test in \cref{sec:analysis setup} and the results are presented in \cref{sec:analysis results}.

\subsection{Analysis setup}
\label{sec:analysis setup}

To evaluate the performance of our emulator in a realistic setting, we embed it in a synthetic version of the DES-Y1 $3\times2$pt analysis. This framework combines three two-point correlation functions: cosmic shear (the auto-correlation of galaxy shapes), galaxy–galaxy lensing (the cross-correlation between galaxy positions and shapes), and galaxy clustering (the auto-correlation of galaxy positions). Together, these observables jointly constrain the underlying matter distribution and the galaxy–matter connection. We adopt the DES-Y1 setup as a representative case study due to its well-documented likelihood pipeline, realistic survey characteristics, and relevance to current and future cosmological experiments.

If we consider two probes (either clustering or shear), labelled by $(i,j)$, the angular power spectrum $C_{\ell}^{ij}$ is given under the Limber approximation as \cite{LoVerde_2008}
\begin{equation}
    C_{\ell}^{ij} \approx \left( \ell + \frac{1}{2} \right)^{\mu_i + \mu_j}
    \int_{a_{ij}}^1 \frac{{\rm d} a}{c^2 \chi^2} \frac{{\rm d} \chi}{{\rm d}a}
    W_i (\chi) W_j(\chi)
    P \left(
    k = \frac{\ell + 1/2}{\chi}, z
    \right),
\end{equation}
where $\mu_i=0$ for galaxy clustering and $\mu_i=-2$ for weak lensing, and the integration range (specified by $a_{ij}$) is determined by the maximum redshift considered.
The number density of tracers in a given tomographic redshift bin is given by $W_i(z)$ which, for galaxy number counts is assumed to be
\begin{equation}
    W^{\rm g}_i (z) = n_i (z) b(z) H(z),
\end{equation}
where $n_i(z)$ is the distribution of redshifts for the sample and $b(z)$ is the bias function.
For weak lensing, the form of $W_i(z)$ is
\begin{equation}
    W^{\rm \kappa}_i (z) = 
    \frac{3 H_0^2 \Omega_{\rm m}}{2 c^2}
    \left( \frac{(\ell + 2)!}{(\ell - 2)!} \right)^{1/2}
    (1 + z) \chi(z)
    \int_z^\infty p_i(z^\prime) \frac{\chi(z^\prime) - \chi(z)}{\chi(z^\prime)} {\rm d}z^\prime
    + W_{\rm IA}(z),
\end{equation}
where $p_i(z) = n_i(z) / \bar{n}_i$ for $\bar{n}_i = \int_0^{z_{\rm max}} n_i(z) \, {\rm d} z$.
We model the intrinsic alignment term, $K_{\rm IA}(z)$, as \cite{Joachimi_2011}
\begin{equation}
    W_{\rm IA}(z) = 
     \left( \frac{(\ell + 2)!}{(\ell - 2)!} \right)^{1/2}
     p_i(z) b(z) H(z) \frac{\Omega_{\rm m}}{D(z)} A_{\text{IA}}(z),
\end{equation}
where
\begin{equation}
    A_{\text{IA}}(z) = A_{\text{IA}} \left( \frac{1+z}{1.62} \right)^{\eta_{\text{IA}}},
\end{equation}
gives the intrinsic alignment amplitude, and is parametrised by the free parameters $A_{\text{IA}}$ and $\eta_{\text{IA}}$.

Our approach in this section broadly follows the analyses of \cite{Krause_2017,Abbott_2018,Campagne_2023}. 
We use the $n_i(z)$ relevant for the four source and five lens samples of DES-Y1. We assume that these may be slightly mis-calibrated, and thus add in one free parameter for each sample, such that
\begin{equation}
    n_i(z) = n_i^{\rm PZ} \left( z - \Delta z^i \right),
\end{equation}
where $n_i^{\rm PZ}$ is the initial estimate of $n_i(z)$.
We consider a linear galaxy bias $b_i(z)=b_i$, which is also a free parameter, and a multiplicative bias for the weak lensing measurements, $\{m_i\}$. We give the priors for each of these parameters used in the inference, as well as for the cosmological parameters, in \cref{tab:mcmc_prior}.
This corresponds to 25 parameters to infer (5 cosmological and 20 `nuisance' parameters), representing a moderately high-dimensional inference problem.

\begin{table}
    \centering
    \begin{tabular}{llll}
        \textbf{Parameter} & \textbf{Prior} & \textbf{Parameter} & \textbf{Prior} \\
        \hline
        \hline
        \multicolumn{2}{l}{\textbf{Cosmology}}  & \multicolumn{2}{l}{\textbf{Lens photo-$z$ shift}} \\
        $\Omega_{\rm m}$ & Uniform (0.1, 0.5) & $\Delta z_1^l$ & Normal (0, 0.007) \\
        $10^{10}A_{\rm s}$ & Uniform (5, 50) & $\Delta z_2^l$ & Normal (0, 0.007) \\
        $n_{\rm s}$ & Uniform (0.8, 1.2) & $\Delta z_3^l$ & Normal (0, 0.006) \\
        $\Omega_{\rm b}$ & Uniform (0.03, 0.07) & $\Delta z_4^l$ & Normal (0, 0.010) \\
        $h$ & Uniform (0.5, 0.9) & $\Delta z_5^l$ & Normal (0, 0.010) \\
        \hline
        \multicolumn{2}{l}{\textbf{Lens Galaxy Bias}} & \multicolumn{2}{l}{\textbf{Shear calibration }} \\
        $b_i\ (1 \leq i \leq 5)$ & Uniform (0.8, 3.0) & $m_i\ (1 \leq i \leq 4)$ & Normal (0, 0.023) \\
        \hline
        \multicolumn{2}{l}{\textbf{Source photo-$z$ shift}} & \multicolumn{2}{l}{\textbf{Intrinsic Alignment}}  \\
        $\Delta z_1^s$ & Normal (0, 0.016) & $A_{\text{IA}}$ & Uniform ($-5$, $5$)\\
        $\Delta z_2^s$ & Normal (0, 0.013) & $\eta_{\text{IA}}$ & Uniform ($-5$, $5$)\\
        $\Delta z_3^s$ & Normal (0, 0.011) \\
        $\Delta z_4^s$ & Normal (0, 0.022) \\
    \end{tabular}
    \caption{Priors used for the mock cosmological analysis. For `Uniform' priors, the arguments give the minimum and maximum allowed values. For `Normal' priors, the prior is a Gaussian distribution with a mean and standard deviation given by the first and second arguments, respectively.}
    \label{tab:mcmc_prior}
\end{table}

For our ``true'' inference which we compare against, we use the DES-Y1 likelihood implemented in \cosmosis\footnote{\url{https://cosmosis.readthedocs.io/en/latest/} } \cite{CosmoSIS_2015}. For both our mock data and inference, we choose to generate data according to the Takahashi et al. 2012 \cite{Takahashi_2012} implementation of \halofit, which is computed using \camb. 
We use the \textsc{multinest} \cite{Feroz_2008,Feroz_2009,Feroz_2019} sampler with 500 live points and allowing a maximum number of 50000 iterations. We terminate the sampling when our evidence estimate reaches an error of 0.1 and we do not use the constant efficiency mode. The target efficiency for the importance nested sampling is 0.3. Otherwise, we use the default settings of \cosmosis. We use a single node with 128 CPU cores to perform our inference.

We compare the inference results from \cosmosis{} with an implementation of this likelihood written in \textsc{jax}, for which we use many of the functionalities of \jaxcosmo{} \cite{Campagne_2023}.
To enable a fair comparison, we modify \jaxcosmo's photometric redshift implementation from a kernel density estimator to a cubic spline, which is what is used by \textsc{cosmosis}. We found that the difference between these two interpolation methods was much greater (up to a few percent) than the difference between the spectra produced by the symbolic emulators and that from using \camb.

Utilising gradients to improve efficiency, we choose to sample with a No-U-Turn Sampler (NUTS) \cite{Hoffman_2011}, as implemented in \textsc{numpyro} \cite{Bingham_2018,Phan_2019}.
We run 16 parallel chains, each with 1000
warm-up steps and 10,000 samples,
using a single Nvidia RTX 2080 GPU. For the \syren{} runs we are able to use a single node of 8 CPU cores with a combined memory of 30.3 GB. This is insufficient memory when we do not use the symbolic emulators, so we use two nodes for those cases (but with a single GPU still).

To compare our run time against that using a numerical emulator, we also run our inference using \textsc{cosmopower-jax} \cite{Piras_2023}. This emulator predicts a different approximation than \halofit{} and thus we do not compare its results at the level of posteriors to that obtained with \cosmosis, but we use it as a useful timing comparison.

\subsection{Results}
\label{sec:analysis results}

The primary goal of an emulator (whether numerical or symbolic) is for increased computational speed. In our inference, there are two primary ways in which our symbolic emulators help in this regard. Firstly, a single likelihood evaluation is much faster since evaluating $P(k)$ is much slower when using \camb{} than our emulator. Second, given the differentiable nature of our emulators, we can gain sampling efficiency through the use of gradients. As such, we find that our inference takes far less time than using \cosmosis. Using the settings outlined in the preceding section, the `exact' inference required approximately 72 hours of run time on 128 CPU cores, whereas our inference only took 
1.2 hours
on a single GPU: approximately 
60
times faster.

It is unsurprising that using our emulator is much faster than the \cosmosis{} inference. However, an interesting result can be seen by comparing our emulator to \jaxcosmo{} and \cosmopower{}. For \jaxcosmo{} we adapt the analysis of \cite{Campagne_2023} for a $\Lambda$CDM cosmology, which notably involves applying \halofit{} to the \eh{} approximation. For the \cosmopower{} result, we replace \eh+\halofit{} with the \cosmopower{} emulator. 
Upon doing this, we find that \cosmopower{} requires 
3.5
hours of runtime, whereas \jaxcosmo{} requires 
9.3
hours. 
We therefore find that using all our emulators results in a speed up of at least a factor of 3. We find that this increase in speed is dominated by the approximations for the comoving distance and growth factor, since we find more modest increases in speed of around 20\% if we only replaced \cosmopower{} by our power spectrum emulator. Since these are our most accurate emulators (they are better than 0.001\% an 0.05\% accurate, respectively), even if one wished to use a neural network for estimating $P(k)$, using our approximations for these quantities would still result in a significant increase in speed since one no longer must numerically solve an integral or differential equation.
We note that neither the neural network approach nor our symbolic emulators have been optimised for speed. As such, either or both of these emulation strategies could be made faster. Nonetheless, for currently available symbolic models and neural networks, we find the former to be faster.

One of the motivations for producing symbolic emulators for various parts of the pipeline, rather than just the linear power spectrum, is to reduce memory usage. In particular, obtaining the \halofit{} variables involves computing integrals and calculating the growth factor requires solving a differentiable equation.
To demonstrate the advantage of these emulators, we compare the memory requirements for computing the mean of the likelihood at the true parameters using a \jaxcosmo{} approach, compared to relying solely on our symbolic emulators. We find that the memory needed to evaluate the mean is comparable, however evaluating the gradient of the \jaxcosmo{} pipeline with respect to the input parameters requires 50\% more memory than for our symbolic emulators. This additional saving in memory could be exploited in a computational analysis by running more MCMC chains in parallel per GPU.

\begin{figure}[htbp]
    \centering
    \begin{subfigure}[t]{0.6\textwidth}
        \centering
        \includegraphics[width=\linewidth]{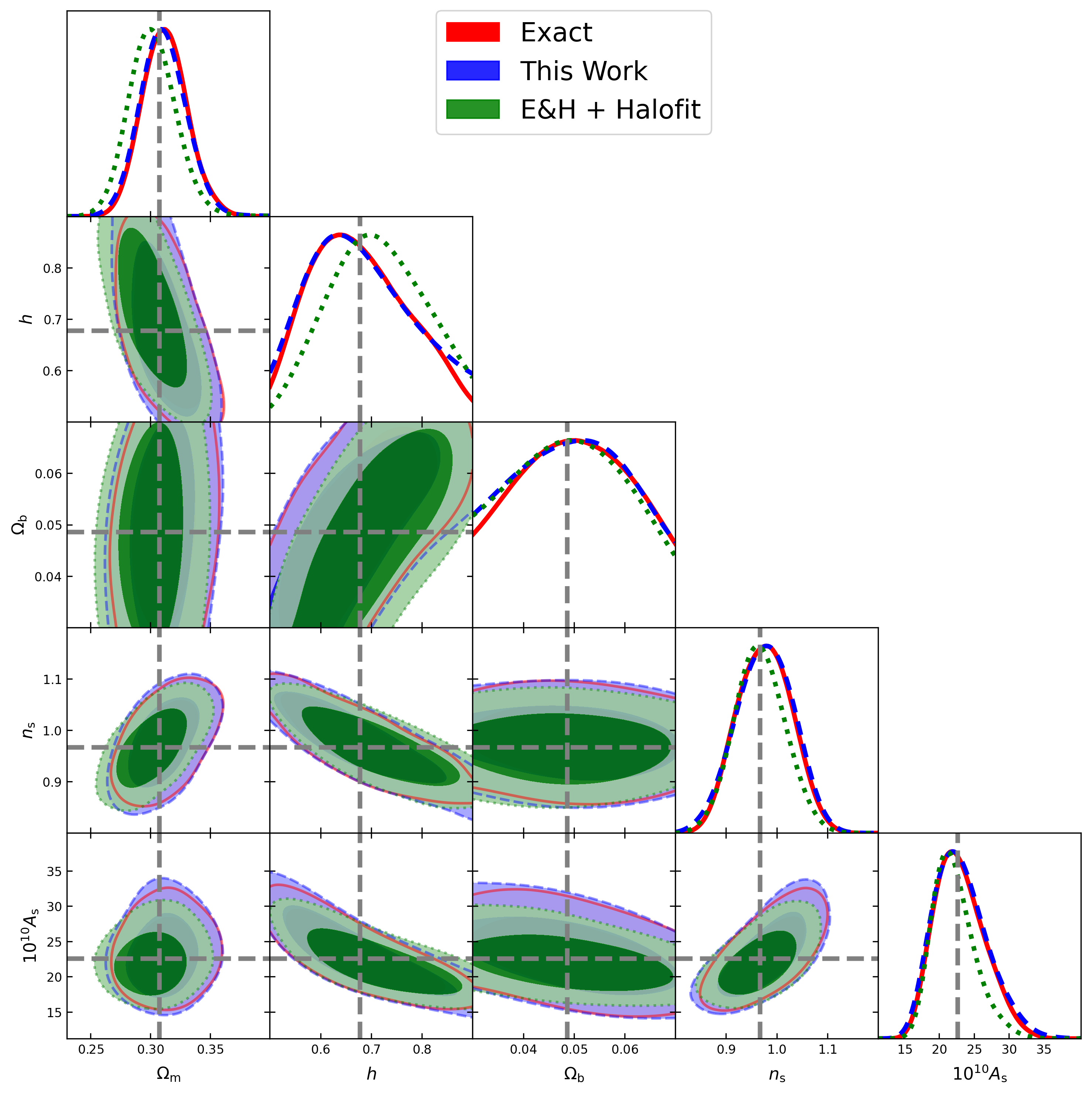}
        \caption{}
        \label{fig:corner_cosmo}
    \end{subfigure}
    \hfill
    \begin{subfigure}[t]{0.38\textwidth}
        \centering
        \includegraphics[width=\linewidth]{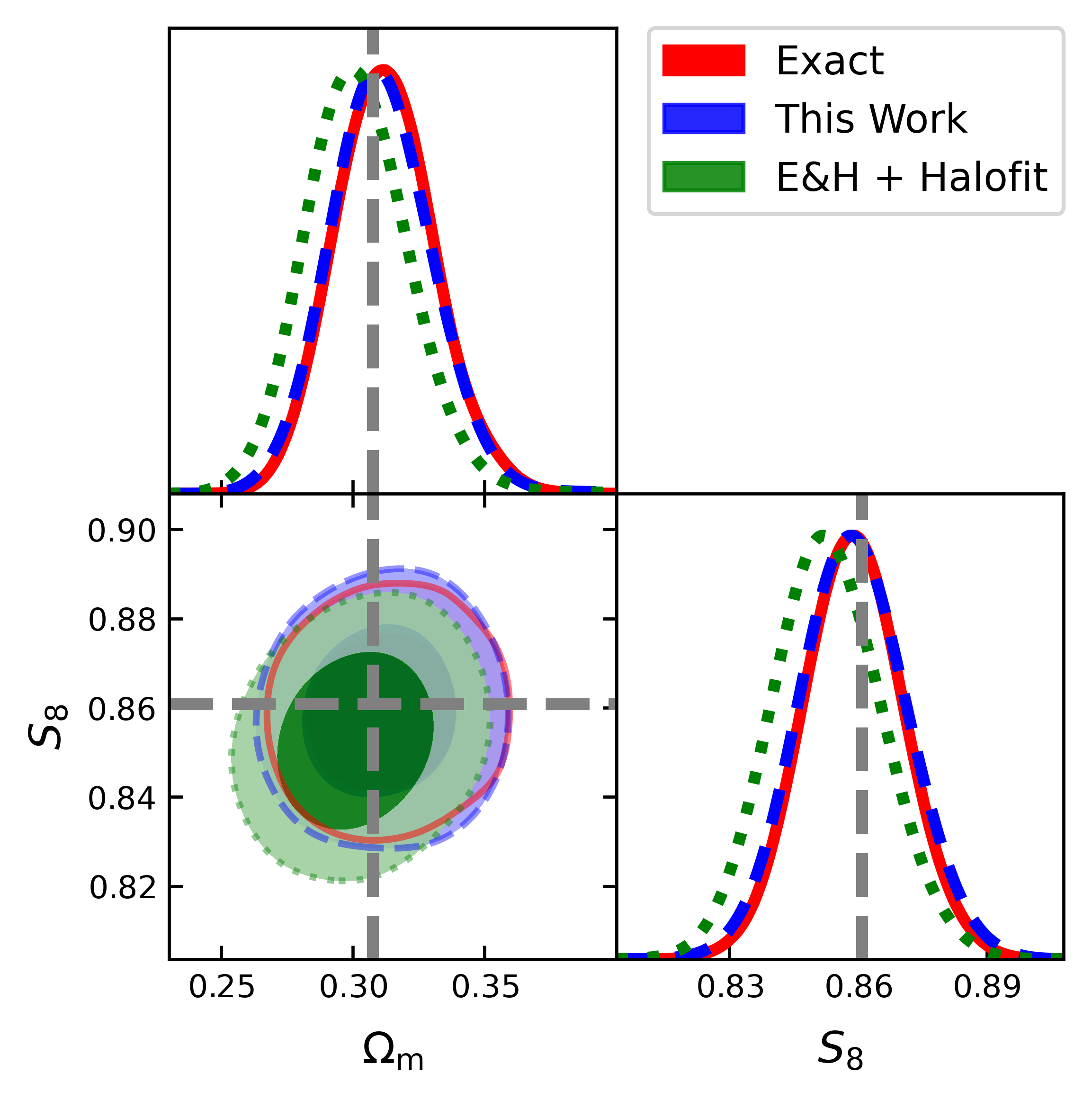}
        \caption{}
        \label{fig:corner_S8}
    \end{subfigure}
    \caption{One- and two-dimensional posterior distributions of the cosmological parameters for our mock cosmological DES-Y1 like analysis. The red contours are obtained using an `exact' model (\camb) and are compared to those obtained using the symbolic approximations produced in this work (blue contours), and to the combination of \eh{} and \halofit{}, as implemented in \jaxcosmo{} (green contours). The grey dashed lines indicate the true parameters. In the left panel we consider the sampled cosmological parameters (which all have uniform priors) whereas in the right panel we also consider a derived parameter, $S_8$. The high level of consistency between our contours and the exact model indicate that the emulators are sufficiently accurate for such an analysis. This is not true for the \eh+\halofit{} model.
    }
    \label{fig:corner_plots}
\end{figure}

Of course, accelerating an inference is only worth doing if the results do not change after doing so. To assess this, in \cref{fig:corner_plots} we plot the one- and two-dimensional posterior distributions for inferred cosmological parameters, plotted using \textsc{getdist} \cite{Lewis_2019}. We see that the distributions inferred using \cosmosis{} and our symbolic emulators are very similar, in particular for the parameters that are well-constrained in these analyses: $\Omega_{\rm m}$, $n_{\rm s}$ and $A_{\rm s}$. There is a small difference between the posteriors of $h$ and $\Omega_{\rm b}$, however these are relatively unconstrained by such an analysis, and thus this is not problematic.
These results are in contrast to the combination of \eh{} and \halofit, with significant shifts in the peaks of the posterior of $\Omega_{\rm m}$ and $n_{\rm s}$, and an overly-confident inference of $A_{\rm s}$.

It is common to not only consider the cosmological parameters as given in \cref{tab:parameter_prior}, but also consider the derived parameter $S_8$, defined to be
\begin{equation}
    S_8 \equiv \sigma_8 \sqrt{\frac{\Omega_{\rm m}}{0.3}},
\end{equation}
since $S_8$ and $\Omega_{\rm m}$ represent two approximately independent parameters which are constrained in such an analysis.
Hence, in \cref{fig:corner_plots} we also show the posteriors in the $\Omega_{\rm m}$-$S_8$ plane, as is conventional. Again, we see that the differences from the truth for the emulators developed in this work are negligible, and thus the symbolic emulators are sufficiently accurate to be used to infer these parameters in such an analysis.
Again, this is not true for \eh+\halofit, with an under-estimation of both $\Omega_{\rm m}$ and $S_8$.

\begin{figure*}
    \centering
    \includegraphics[width=\textwidth]{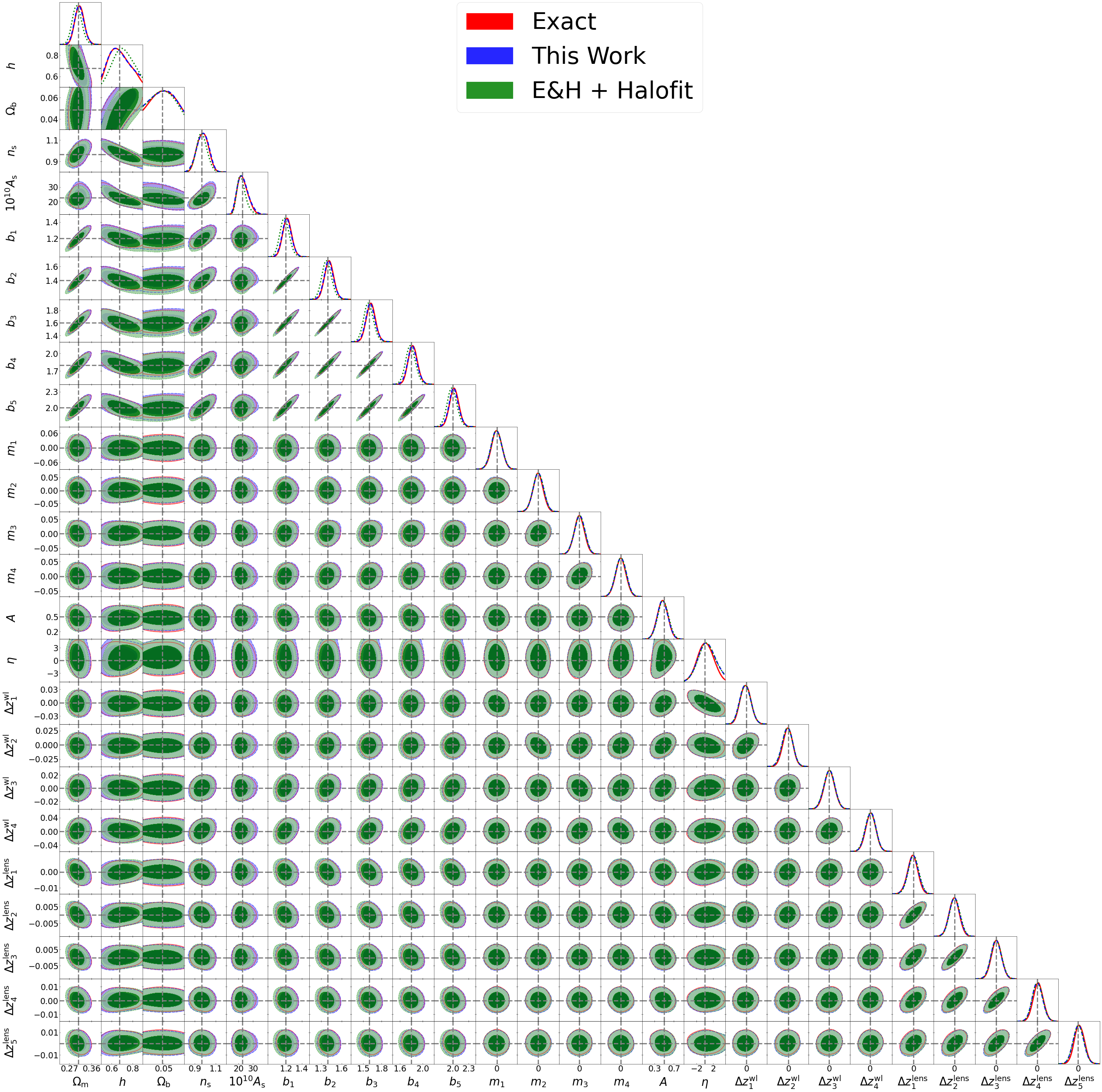}
    \caption{Marginalised posterior distributions of all parameters considered in our mock DES-Y1 analysis. The `exact' model is computed using \camb{}, and we see that its contours (red) are very similar to those obtained using our symbolic emulators (blue), demonstrating that the emulators are sufficiently accurate for such an analysis. This is in contrast to applying \halofit{} to the \eh{} approximation (green contours), for which significant discrepancies are seen.
    \label{fig:corner_all}
    }
\end{figure*}

During our inference, as well as cosmological parameters, we also sampled parameters describing intrinsic alignments, bias, shear calibration errors, and errors in the photometric redshift distributions. To determine whether there are any biases in these parameters, in \cref{fig:corner_all} we plot the two-dimensional marginalised posterior distribution of all parameters sampled in the MCMC. As with the cosmological parameters, we find that the two sets of contours agree well, further demonstrating that the inference with symbolic emulators is accurate. Once again, we see that this is not the case if one used the \halofit{} formalism but with the linear matter power spectrum from \eh{}. Hence, the corrections to \eh{} obtained in this work are necessary to achieve accurate posteriors.

Therefore, we conclude that the symbolic emulators developed in this work are sufficiently accurate to lead to accurate cosmological inference, but this is not true if one uses \eh{} for the linear part of the power spectrum.

\section{Conclusion}
\label{sec:conclusion}

Symbolic emulators offer a powerful alternative to traditional numerical surrogates for cosmological inference. By learning fast, interpretable approximations to expensive forward models, they enable efficient and transparent exploration of cosmological parameter space. In this work, we have extended the validity of symbolic emulators to broader parameter ranges relevant for current analyses, demonstrating that these models retain the accuracy needed for unbiased cosmological inference.
We also introduced a new emulator for the hypergeometic function needed to compute the comoving distance in a $\Lambda$CDM cosmology, which was accurate to better than one part in $10^5$ for all cosmological applications.
The extrapolation behaviour of this emulator is particularly notable, as it become more accurate as one extrapolates beyond the range of the training data (in the direction relevant for cosmology).

We embedded these extended symbolic emulators within a DES Y1-like $3\times2$pt likelihood pipeline and found that the resulting parameter constraints are consistent with those obtained using \camb. Crucially, the symbolic models enable over an order-of-magnitude improvement in evaluation speed, making them especially attractive for applications where computational resources are a limiting factor.
We find that our pipeline with all symbolic emulators is both faster and more memory efficient than if using a combination of numerical emulators for the power spectrum and solving for the growth factor numerically.

In this work, we focused on a Stage-III cosmological analysis as a first step towards deploying symbolic emulators in Stage-IV surveys. The errors on the emulators produced in the this work are sub-dominant compared to the errors on the quantities they are emulating (\halofit{} is insufficiently accurate for Stage-IV surveys). Other symbolic emulators in the \syren{} family overcome this problem by fitting more accurate power spectra \cite{Bartlett_2024_syren,Sui_2024} but with narrower prior ranges than considered here. We leave the extension of their predictions to wider priors and testing their use within Stage-IV likelihoods to future work.

As cosmological datasets continue to grow in precision and volume, the ability to perform fast and accurate inference will only become more essential. The results of this work show that symbolic emulators are not only a viable replacement for numerical ones, but may in fact be preferable in scenarios demanding interpretability, efficiency, and portability. We expect symbolic techniques to play an increasingly central role in the future of cosmological data analysis.

\vskip6pt

\ack{We thank Harry Desmond, Pedro Ferreira, Lukas Kammerer, Gabriel Kronberger, Constantinos Skordis and Benjamin Wandelt for useful comments and suggestions.
The authors are supported by the Simons Collaboration on ``Learning the Universe.''
DJB is supported by Schmidt Sciences through The Eric and Wendy Schmidt AI in Science Fellowship.
This work has made use of the Infinity Cluster hosted by Institut d'Astrophysique de Paris; we thank Stephane Rouberol for running this cluster smoothly for us.
We thank Jonathan Patterson for smoothly running the Glamdring Cluster hosted by the University of Oxford, where some of the data processing was performed.
For the purposes of open access, the authors have applied a Creative Commons Attribution (CC BY) licence to any Author Accepted Manuscript version arising.
The data underlying this article will be shared on reasonable request to the corresponding authors.
We provide implementations of our emulators at \url{https://github.com/DeaglanBartlett/symbolic_pofk}.
}

\begin{appendix}

\section{Symbolic Approximations for the Linear Matter Power Spectrum}
\label{app:pk_lin}

In this section, we give the symbolic expressions for the various terms given in \cref{eq:pk_lin_fit}.
After some manipulations to merge superfluous constants and rounding some which are close to 0 or 1, we obtain the expressions 
\begin{equation}
    \begin{split}
        & f_1 = e_0 \Bigg[ k +
        e_1 \left(1 - e_2 \Omega_{\rm m}^{-e_3}k \right)\\
        & \left( - \Omega_{\rm m} 
        + \frac{e_4 \Omega_{\rm b} + e_5 k}{\sqrt{e_6 + \left( \Omega_{\rm b} + e_7 k\right)^2}}
        - \frac{e_8 \left(e_9 \Omega_{\rm m}\right)^{-e_{10} \Omega_{\rm b}} k}{\sqrt{\left(  e_{11} + h^2\right) \left(e_{12} k^2 + \Omega_{\rm m}^2 \right)}}
        + \frac{e_{13} \left( e_{14} \Omega_{\rm m} \right)^{-e_{15} \Omega_{\rm b} + e_{16} k} k}{\sqrt{e_{17} + \left( e_{18} \Omega_{\rm m}\right)^{e_{19} h} \left( e_{20} \Omega_{\rm m} - k\right)^2}}
        \right) \\
        &
        \left(e_{21} + \left(e_{22} + \frac{\left( e_{23} - e_{24} \Omega_{\rm b} - k\right)^2}{e_{25} + \left(k - e_{26} \Omega_{\rm b} \right)^2} \right)^{-1} \right)^{-\frac{1}{2}}
        \left( e_{27} + \left( e_{28} \Omega_{\rm b} + e_{29} - k \right)^2 \right)^{-\frac{1}{2}}
        \Bigg],
    \end{split}
\end{equation}
\begin{equation}
    f_2 = \left( 1 + 
    \left( 
    \left(e_{30} \Omega_{\rm m}\right)^{-e_{31} \Omega_{\rm b}} \frac{\cos(e_{32} k) - \cos(e_{33} k)}{\sqrt{1 + \cos^2(e_{34} k)}}
    + e_{35} \frac{\left( e_{36} \Omega_{\rm m}\right)^{e_{37}h} \left(e_{38}k\right)^{-e_{39}\Omega_{\rm b}}}{\sqrt{e_{40} + h^2}}k^2
    \right)^2
    \right)^{-\frac{1}{2}},
\end{equation}
\begin{equation}
    f_3 = 
    \left(\frac{e_{41} \left( e_{42} \Omega_{\rm m}\right)^{e_{43} h}}{\sqrt{e_{44} + h^2}}
    - e_{45} \left( e_{46} \Omega_{\rm m}\right)^{e_{47} \Omega_{\rm b} h - e_{48} h^2}\right) k
    + \frac{e_{49}}{\sqrt{e_{50} + \left( \Omega_{\rm m} e_{51} - k \right)^2}},
\end{equation}
\begin{equation}
    \begin{split}
        f_4 &= 
        \left( 
        \frac{e_{52} \left( e_{53} \Omega_{\rm m}\right)^{e_{54} h} k}{\sqrt{e_{55} + \left( e_{56} \Omega_{\rm b} - k\right)^2}\left(\Omega_{\rm b} + e_{57} k \right)}
        + \frac{e_{58}k + e_{59}}{\sqrt{e_{60} + \left(e_{61} \Omega_{\rm m} \right)^{-e_{62}h} k^2}}
        - e_{63}
        - \frac{e_{64}}{\sqrt{e_{65} + k^2}}
        \right) \\
        & 
        \left( 1 + e_{66} \left( - \Omega_{\rm m} + \frac{e_{67} \Omega_{\rm b} + e_{68} \frac{\Omega_{\rm m}}{\Omega_{\rm b}} \left( e_{69} \Omega_{\rm m}\right)^{-e_{70} h} + e_{71} k}{\sqrt{e_{72} + \left( k - e_{73} \right)^2}}\right)^2\right)^{-\frac{1}{2}},
    \end{split}
\end{equation}
where the parameters $\{e_i\}$ are listed in \cref{tab:pk_lin_par}.

\begin{table}
  \centering
  \begin{tabular}{c c | c c | c c | c c}
 Name & Value & Name & Value & Name & Value & Name & Value \\
    \hline\hline
    $e_{0}$ & 0.013924 & $e_{1}$ & 5.1771 & $e_{2}$ & 20.636 & $e_{3}$ & 0.49092 \\
    $e_{4}$ & 3.4224 & $e_{5}$ & 0.35621 & $e_{6}$ & 0.016739 & $e_{7}$ & 0.18401 \\
    $e_{8}$ & 0.58832 & $e_{9}$ & 5.108 & $e_{10}$ & 10.783 & $e_{11}$ & 0.0043879 \\
    $e_{12}$ & 0.11547 & $e_{13}$ & 1.3869 & $e_{14}$ & 2.9301 & $e_{15}$ & 32.014 \\
    $e_{16}$ & 0.002192 & $e_{17}$ & 0.002926 & $e_{18}$ & 316.2 & $e_{19}$ & 1.2158 \\
    $e_{20}$ & 0.095822 & $e_{21}$ & 0.074921 & $e_{22}$ & 0.0067841 & $e_{23}$ & 0.0093912 \\
    $e_{24}$ & 0.022678 & $e_{25}$ & $9.3762 \times 10^{-4}$ & $e_{26}$ & 0.64834 & $e_{27}$ & 0.006882 \\
    $e_{28}$ & 0.6518 & $e_{29}$ & 0.11855 & $e_{30}$ & 2.6863 & $e_{31}$ & 42.261 \\
    $e_{32}$ & 121.6 & $e_{33}$ & 123.21 & $e_{34}$ & 14.51 & $e_{35}$ & 1378.1 \\
    $e_{36}$ & 22.375 & $e_{37}$ & 1.0369 & $e_{38}$ & 75.439 & $e_{39}$ & 3.6282 \\
    $e_{40}$ & 0.15279 & $e_{41}$ & 41.884 & $e_{42}$ & 1052.3 & $e_{43}$ & 0.22687 \\
    $e_{44}$ & 0.022003 & $e_{45}$ & 236.72 & $e_{46}$ & 0.16852 & $e_{47}$ & 0.55815 \\
    $e_{48}$ & 0.040376 & $e_{49}$ & 0.4352 & $e_{50}$ & 0.011574 & $e_{51}$ & 0.043423 \\
    $e_{52}$ & 0.001834 & $e_{53}$ & 205.7 & $e_{54}$ & 0.26165 & $e_{55}$ & 0.0073544 \\
    $e_{56}$ & 0.99135 & $e_{57}$ & 0.18444 & $e_{58}$ & 0.089257 & $e_{59}$ & 0.0074827 \\
    $e_{60}$ & $7.7431 \times 10^{-5}$ & $e_{61}$ & 1.1788 & $e_{62}$ & 0.44205 & $e_{63}$ & 0.098723 \\
    $e_{64}$ & 0.01075 & $e_{65}$ & 0.007492 & $e_{66}$ & 0.82882 & $e_{67}$ & 1.1882 \\
    $e_{68}$ & 383.17 & $e_{69}$ & 397.51 & $e_{70}$ & 4.4661 & $e_{71}$ & 0.81134 \\
    $e_{72}$ & 0.0052959 & $e_{73}$ & 0.15939 \\
  \end{tabular}
  \caption{Best-fit parameters for the linear matter power spectrum emulator (\cref{eq:pk_lin_fit}).}
  \label{tab:pk_lin_par}
\end{table}

\section{Symbolic Approximations for Halofit Variables}
\label{app:halofit}

In this section we give our chosen analytic approximations to the halofit variables defined in \cref{eq:ksigma definition,eq:neff C definition}. For the nonlinear scale, $k_\sigma$, we choose
\begin{equation}
    \label{eq:ksigma_fit}
    \begin{split}
        &\log k_\sigma =
        -\psi_0 \sigma_8 - \left( \psi_1 h \right)^{\psi_2 z \left( \psi_3 \Omega_{\rm m} \right)^{\psi_4 \sigma_8 -\psi_5 \sqrt{z}}}
        + \frac{\psi_6}{h} \left( \psi_7 z\right)^{\psi_8 n_{\rm s} + \psi_9 \sigma_8 - \psi_{10} z}\\
        & + \Bigg[
        - \psi_{11} z \left( - \psi_{12} \Omega_{\rm b} + \left( \psi_{13} \Omega_{\rm m} + \log(\psi_{14} n_{\rm s}) \right)^{-\psi_{15} z (\psi_{16} z)^{-\psi_{17} h}}\right) \\
        & - 
        \left( \psi_{18} \sigma_8 (\psi_{19} \Omega_{\rm m})^{-\psi_{20} z} - (\psi_{21} n_{\rm s})^{\left( - \psi_{22}n_{\rm s} - \psi_{23} z\right) (\psi_{24} h)^{\psi_{25} z}} \right) \\
        & \times \left( - \psi_{26} h + \psi_{27} n_{\rm s} + \left( -\psi_{28} \Omega_{\rm b} + \psi_{29} \Omega_{\rm m} \right)^{-\psi_{30} \Omega_{\rm b} - \psi_{31} \sigma_8} \right)
        \Bigg] \\
        & \times 
        \left[ \psi_{32} \Omega_{\rm m} + \psi_{33} \sigma_8 - \psi_{34} + (\psi_{35} n_{\rm s})^{\psi_{36} n_{\rm s} + \psi_{37} z} + \left( - \psi_{38} \Omega_{\rm b} + \psi_{39} \Omega_{\rm m} \right)^{\psi_{40} \sigma_8} + (\psi_{41} n_{\rm s})^{\psi_{42} \sigma_8 (\psi_{43} h)^{-\psi_{44} z}}\right]^{-1},
    \end{split}
\end{equation}
where the parameters $\{\psi_i\}$ are tabulated in \cref{tab:ksigma_par}.

Our approximation to the effective slope, $n_{\rm eff}$, is
\begin{equation}
    \label{eq:neff_fit}
    \begin{split}
         n_{\rm eff} & \approx
            \left(\chi_0 \sqrt{n_{\rm s}} - \chi_1 \right) 
            \left( \chi_2 n_{\rm s} + \chi_3 \sigma_8 + \chi_4 z - \chi_5 \right)^{\chi_6 \sigma_8}\\
            &  \times
            \left( \chi_7 \sqrt{\chi_8 \Omega_{\rm m} - \Omega_{\rm b}} - \frac{\chi_9}{h} \left(\chi_{10} \sqrt{n_{\rm s}} + \left( \chi_{11} \Omega_{\rm m} \right)^{-\chi_{12} z - \chi_{13}} \right)^{-\chi_{14} \Omega_{\rm m}}\right) \\
            & \times \left(\chi_{15} h + \chi_{16} \sigma_8 + \left(\chi_{17} z\right)^{-\chi_{18} z} \right),
    \end{split}
\end{equation}
where the optimised values of $\{\chi_i\}$ are given in \cref{tab:neff_par}.

Finally, our approximation to the curvature variable, $C$, is
\begin{equation}
    \label{eq:C_fit}
    \begin{split}
        C & \approx
        \left(\varphi_{0} z\right)^{- \varphi_{1} z} \\
        & \times \left(\varphi_2 \Omega_{\rm m} - \varphi_3 \Omega_{\rm b} + \varphi_{4} z\right)^{\varphi_{5} + \left(\varphi_6 \Omega_{\rm m}\right)^{- \varphi_{7} h + \varphi_{8} n_{s} - \varphi_{9} \sigma_{8}} - \left(\varphi_{10} z + \varphi_{11} n_{s}\right)^{- \varphi_{12} \Omega_{m} - \varphi_{13} h - \varphi_{14} \sigma_{8} - \varphi_{15} z}} \\
        & \times \left(
        - \varphi_{16} h - \varphi_{17} \sigma_{8} + \left(\varphi_{18} \sigma_{8}\right)^{\left(\varphi_{19} \Omega_{m} - \varphi_{20} \Omega_{b} + \varphi_{21}\right)^{- \varphi_{22} h} \left(\varphi_{23} \Omega_{m} + \varphi_{24} h + \varphi_{25} n_{\rm s} + \varphi_{26} \sigma_{8} - \varphi_{27} z\right)} \right. \\
        & \left. - \left(- \varphi_{28} \Omega_{\rm m} + \varphi_{29} n_{\rm s} + \varphi_{30} z\right)^{\varphi_{31} h + \varphi_{32} n_{\rm s} + \varphi_{33} \sigma_{8}}
        \right),
    \end{split}
\end{equation}
with the optimised coefficients $\{\varphi_i\}$ listed in \cref{tab:C_par}.

\begin{table}
  \centering
  \begin{tabular}{c c | c c | c c | c c | c c}
 Name & Value & Name & Value & Name & Value & Name & Value & Name & Value \\
    \hline\hline
    $\psi_{0}$ & 0.3458 & $\psi_{1}$ & 0.01477 & $\psi_{2}$ & 0.0825 & $\psi_{3}$ & 4.642 & $\psi_{4}$ & 0.4738 \\
    $\psi_{5}$ & 0.3847 & $\psi_{6}$ & 2.005 & $\psi_{7}$ & 0.02206 & $\psi_{8}$ & 0.2958 & $\psi_{9}$ & 0.4962 \\
    $\psi_{10}$ & 0.03355 & $\psi_{11}$ & 0.6467 & $\psi_{12}$ & 1.139 & $\psi_{13}$ & 8.498 & $\psi_{14}$ & 4.57 \\
    $\psi_{15}$ & 0.6448 & $\psi_{16}$ & 0.1022 & $\psi_{17}$ & 0.3782 & $\psi_{18}$ & 1.239 & $\psi_{19}$ & 18.27 \\
    $\psi_{20}$ & 0.1043 & $\psi_{21}$ & 0.3435 & $\psi_{22}$ & 0.2178 & $\psi_{23}$ & 0.1644 & $\psi_{24}$ & 0.2413 \\
    $\psi_{25}$ & 0.03482 & $\psi_{26}$ & 0.5142 & $\psi_{27}$ & 0.4983 & $\psi_{28}$ & 1.1 & $\psi_{29}$ & 0.8871 \\
    $\psi_{30}$ & 0.7256 & $\psi_{31}$ & 0.1405 & $\psi_{32}$ & 0.614 & $\psi_{33}$ & 1.027 & $\psi_{34}$ & 3.036 \\
    $\psi_{35}$ & 0.9132 & $\psi_{36}$ & 0.4545 & $\psi_{37}$ & 0.2444 & $\psi_{38}$ & 113.5 & $\psi_{39}$ & 97.35 \\
    $\psi_{40}$ & 0.1534 & $\psi_{41}$ & 0.9639 & $\psi_{42}$ & 1.309 & $\psi_{43}$ & 1.616 & $\psi_{44}$ & 0.3708 \\
  \end{tabular}
  \caption{Best-fit parameters for the emulator of the \halofit{} variable $k_\sigma$ (\cref{eq:ksigma_fit}).}
  \label{tab:ksigma_par}
\end{table}

\begin{table}
  \centering
  \begin{tabular}{c c | c c | c c | c c | c c}
 Name & Value & Name & Value & Name & Value & Name & Value & Name & Value \\
    \hline\hline
    $\chi_{0}$ & 3.3208 & $\chi_{1}$ & 6.3738 & $\chi_{2}$ & 0.2304 & $\chi_{3}$ & 0.1642 & $\chi_{4}$ & 0.064 \\
    $\chi_{5}$ & 0.1461 & $\chi_{6}$ & 0.2171 & $\chi_{7}$ & 0.8835 & $\chi_{8}$ & 0.7457 & $\chi_{9}$ & 0.0537 \\
    $\chi_{10}$ & 0.268 & $\chi_{11}$ & 6.4778 & $\chi_{12}$ & 2.3502 & $\chi_{13}$ & 1.3872 & $\chi_{14}$ & 0.6122 \\
    $\chi_{15}$ & 0.8784 & $\chi_{16}$ & 0.6466 & $\chi_{17}$ & 512.827 & $\chi_{18}$ & 0.0894 \\
  \end{tabular}
  \caption{Best-fit parameters for the emulator of the \halofit{} variable $n_{\rm eff}$ (\cref{eq:neff_fit}).}
  \label{tab:neff_par}
\end{table}

\begin{table}
  \centering
  \begin{tabular}{c c | c c | c c | c c}
 Name & Value & Name & Value & Name & Value & Name & Value \\
    \hline\hline
    $\varphi_{0}$ & 4.917 & $\varphi_{1}$ & 0.04262 & $\varphi_{2}$ & 1461 & $\varphi_{3}$ & 2181 \\
    $\varphi_{4}$ & 11.15 & $\varphi_{5}$ & 0.4784 & $\varphi_{6}$ & 0.09069 & $\varphi_{7}$ & 0.0343 \\
    $\varphi_{8}$ & 0.04317 & $\varphi_{9}$ & 0.0372 & $\varphi_{10}$ & 0.09107 & $\varphi_{11}$ & 0.151 \\
    $\varphi_{12}$ & 0.04674 & $\varphi_{13}$ & 0.04854 & $\varphi_{14}$ & 0.05496 & $\varphi_{15}$ & 0.03631 \\
    $\varphi_{16}$ & 0.1805 & $\varphi_{17}$ & 0.1707 & $\varphi_{18}$ & 0.2315 & $\varphi_{19}$ & 0.4075 \\
    $\varphi_{20}$ & 0.593 & $\varphi_{21}$ & 1.84 & $\varphi_{22}$ & 1.028 & $\varphi_{23}$ & 0.02645 \\
    $\varphi_{24}$ & 0.06507 & $\varphi_{25}$ & 0.06477 & $\varphi_{26}$ & 0.192 & $\varphi_{27}$ & 0.003867 \\
    $\varphi_{28}$ & $5.56 \times 10^{-4}$ & $\varphi_{29}$ & $8.51 \times 10^{-4}$ & $\varphi_{30}$ & $1.77 \times 10^{-4}$ & $\varphi_{31}$ & 0.03328 \\
    $\varphi_{32}$ & 0.04181 & $\varphi_{33}$ & 0.06002 \\
  \end{tabular}
  \caption{Best-fit parameters for the emulator of the \halofit{} variable $C$ (\cref{eq:C_fit}).}
  \label{tab:C_par}
\end{table}

\end{appendix}

\bibliographystyle{RS}
\bibliography{references}

\end{document}

%% file: journals.tex
\newcommand*\aap{A\&A}
\let\astap=\aap
\newcommand*\aapr{A\&A~Rev.}
\newcommand*\aaps{A\&AS}
\newcommand*\actaa{Acta Astron.}
\newcommand*\aj{AJ}
\newcommand*\ao{Appl.~Opt.}
\let\applopt\ao
\newcommand*\apj{ApJ}
\newcommand*\apjl{ApJ}
\let\apjlett\apjl
\newcommand*\apjs{ApJS}
\let\apjsupp\apjs
\newcommand*\aplett{Astrophys.~Lett.}
\newcommand*\apspr{Astrophys.~Space~Phys.~Res.}
\newcommand*\apss{Ap\&SS}
\newcommand*\araa{ARA\&A}
\newcommand*\azh{AZh}
\newcommand*\baas{BAAS}
\newcommand*\bac{Bull. astr. Inst. Czechosl.}
\newcommand*\bain{Bull.~Astron.~Inst.~Netherlands}
\newcommand*\caa{Chinese Astron. Astrophys.}
\newcommand*\cjaa{Chinese J. Astron. Astrophys.}
\newcommand*\fcp{Fund.~Cosmic~Phys.}
\newcommand*\gca{Geochim.~Cosmochim.~Acta}
\newcommand*\grl{Geophys.~Res.~Lett.}
\newcommand*\iaucirc{IAU~Circ.}
\newcommand*\icarus{Icarus}
\newcommand*\jcap{J. Cosmology Astropart. Phys.}
\newcommand*\jcp{J.~Chem.~Phys.}
\newcommand*\jgr{J.~Geophys.~Res.}
\newcommand*\jqsrt{J.~Quant.~Spectr.~Rad.~Transf.}
\newcommand*\jrasc{JRASC}
\newcommand*\memras{MmRAS}
\newcommand*\memsai{Mem.~Soc.~Astron.~Italiana}
\newcommand*\mnras{MNRAS}
\newcommand*\na{New A}
\newcommand*\nar{New A Rev.}
\newcommand*\nat{Nature}
\newcommand*\nphysa{Nucl.~Phys.~A}
\newcommand*\pasa{PASA}
\newcommand*\pasj{PASJ}
\newcommand*\pasp{PASP}
\newcommand*\physrep{Phys.~Rep.}
\newcommand*\physscr{Phys.~Scr}
\newcommand*\planss{Planet.~Space~Sci.}
\newcommand*\pra{Phys.~Rev.~A}
\newcommand*\prb{Phys.~Rev.~B}
\newcommand*\prc{Phys.~Rev.~C}
\newcommand*\prd{Phys.~Rev.~D}
\newcommand*\pre{Phys.~Rev.~E}
\newcommand*\prl{Phys.~Rev.~Lett.}
\newcommand*\procspie{Proc.~SPIE}
\newcommand*\qjras{QJRAS}
\newcommand*\rmxaa{Rev. Mexicana Astron. Astrofis.}
\newcommand*\skytel{S\&T}
\newcommand*\solphys{Sol.~Phys.}
\newcommand*\sovast{Soviet~Ast.}
\newcommand*\ssr{Space~Sci.~Rev.}
\newcommand*\zap{ZAp}